\documentclass[twocolumn,showpacs,preprintnumbers]{revtex4}

\usepackage{txfonts}
\usepackage{amssymb}
\usepackage[section] {placeins}

\usepackage{graphicx}
\usepackage{dcolumn}
\usepackage{bm}
\usepackage{latexsym}
\usepackage{feynmf}     
\usepackage{longtable}
\usepackage{color}

\begin{document}
\def\bsb{{\hbox{\boldmath $\beta$}}}
\def\undersim#1{\setbox9\hbox{${#1}$}{#1}\kern-\wd9\lower
    2.5pt \hbox{\lower\dp9\hbox to \wd9{\hss $_\sim$\hss}}}
\def\brho{{\hbox{\boldmath $\rho$}}}
\def\bbox#1{\hbox{\boldmath${#1}$}}
\def\gtsim{$\raisebox{0.6ex}{$>$}\!\!\!\!\!\raisebox{-0.6ex}{$\sim$}\,\,$}
\def\ltsim{$\raisebox{0.6ex}{$<$}\!\!\!\!\!\raisebox{-0.6ex}{$\sim$}\,\,$}
\def\pt{\bbox{p}_t}
\def\rr{\hbox{\boldmath{$ r $}}}
\def\pp{\hbox{\boldmath{$ p $}}}
\def\pt{\pp_{{}_T}}
\def\yb{{\bar y }}
\def\mt{m_{{}_T}}
\def\zb{{\bar z }}
\def\tb{{\bar t }}
\def\rhoe{ {\rho_{\rm eff}} }
\def\qq{\hbox{\boldmath{$ q $}}}

\title{Pion transverse-momentum spectrum and elliptic anisotropy of partially coherent source}

\author{Peng Ru$^{1,}$\footnote{Electronic address:
pengru@mail.dlut.edu.cn}}
\author{Ghulam Bary$^{1,}$\footnote{Electronic address:
ghulambary@mail.dlut.edu.cn}}
\author{Wei-Ning Zhang$^{1,\,2,}$\footnote{Electronic address:
wnzhang@dlut.edu.cn}}
\affiliation{$^1$School of Physics, Dalian University of Technology, Dalian, Liaoning
116024, China\\
$^2$Department of Physics, Harbin Institute of Technology, Harbin, Heilongjiang 150006,
China}


\begin{abstract}
In this letter, we study the pion momentum distribution of a coherent source and investigate the influences
of coherent emission on the pion transverse-momentum~($p_T$) spectrum and elliptic anisotropy.
With a partially coherent source, constructed by a conventional viscous hydrodynamics model\,
(chaotic part) and a parameterized expanding coherent source model, we reproduce the pion $p_T$
spectrum and elliptic anisotropy coefficient $v_2(p_T)$ in the peripheral Pb-Pb collisions at
$\sqrt{s_{NN}}=2.76$~TeV. It is found that the influences of coherent emission on the pion $p_T$
spectrum and $v_2(p_T)$ are related to the initial size and shape of the coherent source,
largely due to the interference effect. However, the effect of source dynamical evolution on
coherent emission is relatively small.  The results of the partially coherent
source with 33\% coherent emission and 67\% chaotic emission are consistent with
the experimental measurements of the pion $p_T$ spectrum, $v_2(p_T)$, and especially four-pion
Bose-Einstein correlations.
\end{abstract}

\maketitle

\section{Introduction}
\label{introduction}
The particle transverse-momentum spectrum and elliptic anisotropy are important observables
in relativistic heavy-ion collisions\,\cite{{STAR-spe04,PHENIX-spe04,PHOBOS-spe07,
Abelev:2012wca,ALICE-spe13,STAR-v2-05,PHENIX-v2-09,ALICE-v2-15}}.
The transverse-momentum~($p_T$) spectrum can reveal information
about the thermalization and expansion of the particle-emitting sources produced in
such collisions\,\cite{STAR-spe04,PHENIX-spe04,PHOBOS-spe07,Abelev:2012wca,ALICE-spe13,Heinz:2009xj}.
In addition, the azimuthal anisotropy coefficient $v_n$ is related to the source initial anisotropic
pressure gradient\,\cite{{STAR-v2-05,PHENIX-v2-09,ALICE-v2-15,Heinz:2009xj,Ollitrault:1992bk,
Schenke:2011qd,Snellings:2014kwa}}, the source viscosity\,\cite{{Romatschke:2007mq,Song:2007fn,
Song:2009gc,Schenke:2010rr,Shen:2011eg,Bozek:2011ua,Gale:2012rq,Song:2013qma,Shen:2015msa}},
the uncertainty relation of quantum mechanics\,\cite{MolnarWangGreene14}, and even the particle
escape from a spatially asymmetric source\,\cite{HeEdmLin-PLB16}.

Recently, experimentalists in the ALICE Collaboration observed a significant suppression of three- and four-pion
Bose-Einstein correlations in Pb-Pb collisions at $\sqrt{s_{NN}}=2.76$~TeV at the
Large Hadron Collider (LHC) \cite{ALICE-HBT14,ALICE-HBT16}.  This may indicate that there
is a considerable degree of coherent pion emission in relativistic heavy-ion collisions
\cite{{ALICE-HBT14,ALICE-HBT16,Wiedemann:1999qn,Weiner:1999th,Akkelin:2001nd,WongZhang07,
LiuRuZhangWong14,LiuRuZhangWong13,Gangadharan15}}.  In addition to the pion Bose-Einstein condensation
\cite{WongZhang07,LiuRuZhangWong13,LiuRuZhangWong14,Begun:2015ifa}, the pion laser \cite{Pratt93,Csorgo:1997us}, disoriented chiral condensate (DCC)
\cite{Zhang:1996nq,Zhang:1997wz,AmelinoCamelia:1997in,WA98-PLB98},
gluonic condensate \cite{Blaizot:2011xf,Xu:2014ega,Huang:2013lia}, and even the initial-stage
coherent gluon field \cite{Kovner:1995ja,Iancu:2002xk,Schenke:2016lrs} may possibly give rise to
coherent particle emissions in relativistic heavy-ion collisions.  It is meaningful to explore
the effects of coherent emission on the final-state observables.

The effect of Bose-Einstein condensation on the particle velocity distribution has been
observed in ultra-cold atomic gases \cite{Anderson:1995gf,Davis:1995pg}.
The appearance of a condensate leads to a decrease of the velocity distribution width.
Furthermore, since the frequencies $\omega_i$ of the trapping potential (thus, the trapping sizes $a_i \sim\sqrt{1/\omega_i}$
\cite{LiuRuZhangWong13}) are different in the symmetry-axis direction and
the directions perpendicular to the symmetry axis, the two-dimensional velocity
distribution presents an elliptic pattern when the condensate appears
\cite{Anderson:1995gf,Davis:1995pg}.  This is a quantum-mechanical response to the
asymmetric spatial configuration of the condensate.  The investigation of the analogous
anisotropic particle momentum distribution in relativistic heavy-ion collisions
is of great interest for exploring the origin of coherence of a particle-emitting source.

In this work, we study the pion momentum distribution and azimuthal anisotropy of
coherent and chaotic emissions in relativistic heavy-ion collisions.
We investigate the effects of source geometry and expansion on the pion $p_T$ spectrum and
$v_2(p_T)$.  Furthermore, we construct a partially coherent pion source combined with a
hydrodynamical chaotic source and a parameterized coherent source, and compare the results of
the pion $p_T$ spectrum and $v_2(p_T)$ of the partially coherent source with the experimental data
measured in the Pb-Pb collisions at $\sqrt{s_{NN}}=2.76$~TeV at the LHC.
It is found that the influences of coherent emission on the pion transverse-momentum spectrum and
elliptic anisotropy are related to the initial size and shape of the coherent source, mainly
due to the interference effect.
The results of the partially coherent source with 33\% coherent emission and 67\% chaotic
emission are consistent with the experimental measurements of the pion $p_T$ spectrum,
$v_2(p_T)$, and four-pion Bose-Einstein correlations.

This letter is organized as follows.  In Sec. II, we present the formulas of
momentum distributions for the coherent and chaotic pion emissions.  In Sec. III, we study
the effects of source geometry and expansion on the pion $p_T$ spectrum and $v_2(p_T)$.
In Sec. IV, the results of the partially coherent source are presented and compared
with the experimental data at the LHC.  Finally, we give a summary and discussion in Sec. IV.

\section{Pion momentum distribution for coherent and chaotic emissions}
\label{CS}
A well-known purely coherent multi-particle system is the radiation field of a classical
source (current) \cite{Gla-PR1951,Gla-PRL1963,Gla-PR1963,Gla-PR1963a,Gyu-PRC1979}.
In this model, the final state of the pion field produced by a classical source $\rho(x)
=\rho(t, \rr)$ can be written as~\cite{Gyu-PRC1979}
\begin{eqnarray}
|\phi_{\pi}\rangle =e^{-\bar{n}/2}\exp\left( i\! \int \! d^3\!p \,{\cal A}(\pp)\,
a^\dag(\pp) \right)|0\rangle,
\end{eqnarray}
where $a^\dag(\pp)$ is the pion creation operator for momentum $\pp$, ${\cal A}(\pp)$
is an amplitude related to the on-shell ($E^2_p=\pp^2+m_{\pi}^2$) Fourier transform of
$\rho(x)$,
\begin{eqnarray}
{\cal A}(\pp)=A(\pp)\!\int\!d^4\!x\,e^{i(E_p t-\pp\cdot\rr)}\rho(t,\rr)\equiv A(\pp)\,
\tilde{\rho}(\pp),
\label{FTclas}
\end{eqnarray}
\begin{equation}
A(\pp)=A(E_p)=\left[2E_p(2\pi)^3\right]^{-1/2}  ,
\label{cohpoint}
\end{equation}
and
\begin{eqnarray}
\bar{n}=\!\int \!d^3\!p~\left|{\cal A}(\pp)\right|^2
\end{eqnarray}
is the average pion number in the final state.

Considering that the coherent state $|\phi_{\pi}\rangle$ is an eigenstate of the annihilation operators
$a(\pp)$, i.e. $a(\pp)\,|\phi_{\pi}\rangle=i{\cal A}(\pp)\,|\phi_{\pi}\rangle$,
we can write the single-pion momentum distribution as
\begin{eqnarray}
&&P_{\!_C}(\pp)
\equiv\frac{d^3\bar{n}}{d^3p}
={\rm Tr}\left[D_{\pi}\,a^\dag\!(\pp)\,a(\pp)\right] \nonumber\\
&&\hspace*{19mm}=\left|\,{\cal A}(\pp)\,\right|^2=\left|\,A(\pp)\,\tilde\rho(\pp)\,\right|^2,~~~~
\label{P_c}
\end{eqnarray}
where $D_{\pi}=|\phi_{\pi}\rangle\langle\phi_{\pi}|$ is the density matrix of the coherent state.
One can interpret ${\cal A}(\pp)$ as the amplitude for the classical source $\rho(x)$ to emit a
pion with momentum $\pp$.
A special case is that the source is point-like in space-time, namely $\rho(x)=\delta^{(4)}(x)$,
and then one has ${\cal A}(\pp)=A(\pp)$.
Note that $|A(\pp)|^2$ is the function of $E_p$ and is thus azimuthally isotropic in momentum space.
For general source distributions, the total amplitude ${\cal A}(\pp)$ in the form of
Eq.\,(\ref{FTclas}) can be viewed as the coherent superposition of
the sub-amplitudes $A(\pp)\,e^{ip\cdot x}$ at different space-time coordinates,
where the factor $e^{ip\cdot x}$ can be related to the propagation of the pion\,\cite{CYWong_book}.

An important property of the coherent state $|\phi_{\pi}\rangle$ is that the
multi-pion momentum distribution can be factorized into the product of the
single-pion momentum distributions,
\begin{eqnarray}
&&P_{\!_C}(\pp_1,\dots,\pp_m)\!=\!{\rm Tr}\big[D_{\pi}\,a^\dag\!(\pp_1)\cdots
a^\dag\!(\pp_m)\,a(\pp_m)\cdots a(\pp_1)\big]~~~~~\nonumber\\
&&\hspace*{21mm}=\left|\,{\cal A}(\pp_1)\,\right|^2 \cdots\left|\,{\cal A}(\pp_m)\,
\right|^2.
\end{eqnarray}
Owing to this property, there is no Hanbury-Brown-Tiwss\,(HBT) effect present in a coherent state.
In order to distinguish the amplitude and source density of the chaotic pion
emission to be discussed later, we use the denotations $A_{_C}(\pp)$,
$\rho_{\!_C}(x)$, and $\tilde\rho_{\!_C}(\pp)$ for the source of the coherent
state.

For chaotic pion emission, the single-pion momentum distribution can be similarly written
as the absolute square of the total emission amplitude of the chaotic source~\cite{CYWong_book},
\begin{eqnarray}
P_{\!\chi}(\pp)
=\!\bigg|{\sum_x}A_{\chi}(x,\pp)\, e^{ip\cdot x}\bigg|^2
=\bigg|{\sum_x}\,\big|A_{\chi}(x,\pp)\big|\,e^{i\phi_{\chi}(x)} e^{ip\cdot x}\bigg|^2,
\label{chaofull}
\end{eqnarray}
where $\sum$ denotes the summation over the source points of chaotic emission, and
$A_{\chi}(x,\pp)=\left|A_{\chi}(x,\pp)\right|\,e^{i\phi_{\chi}(x)}$ is the amplitude
for a source point at $x$ to emit a pion with momentum $\pp$, with the phase $\phi_{\chi}(x)$
randomly varying with $x$.
Owing to the randomness of $\phi_{\chi}(x)$, the interference terms in the expansions of the
absolute square in Eq.\,(\ref{chaofull}), for pion emissions from different source points, will
tend to cancel each other out and give a negligible contribution.  In the absence of interference
effects, the single-pion momentum distribution in Eq.\,(\ref{chaofull}) can be written as the
sum over the momentum distributions for all the source points,
\begin{eqnarray}
P_{\!\chi}(\pp)=\!\sum_x\left|A_{\chi}(x,\pp)
\right|^2=\!\!\int d^4\!x\,\rho_{\chi}(x)\,\left|A_{\chi}(x,\pp)\right|^2,
\label{P_ch}
\end{eqnarray}
where $\rho_{\chi}(x)$ is the chaotic source distribution.
A main characteristic of the chaotic emission is that each of the source points emits a pion
independently.  In addition, the two- and multi-pion momentum distributions cannot be expressed
as the product of the single-pion momentum distributions, which gives rise to the HBT correlations
\cite{HBT1956,GGLP1960}.

Assuming that all the sub-sources in the chaotic source emit a pion thermally at the same
emission~(or freeze-out) temperature $T$, we have, for a static source,
\begin{equation}
P_{\!\chi}(\pp)=\left|A_{\chi}(\pp)\right|^2 \sim \frac{1}{e^{E_p/T}-1},
\label{P_chS}
\end{equation}
where the source distribution $\rho_{\chi}(x)$ is assumed to be normalized.
This single-pion momentum distribution for a static chaotic source is the function of $E_p$ and $T$, and
is thus azimuthally isotropic.  We can also see that in this case $P_{\!\chi}(\pp)$ is independent
of the source space-time distribution.
However, due to the interferences between the pion emissions at different source points,
the single-pion momentum distribution for the coherent source depends on
the Fourier transform of the source space-time distribution, $\tilde\rho_{\!_C}(\pp)$\,[as in Eq.\,(\ref{P_c})].
Even for a static source, an azimuthally anisotropic $|\,\tilde\rho_{\!_C}(\pp)\,|^2$ will result in
an anisotropic $P_{\!_C}(\pp)$.

Fundamental distinctions between coherent and chaotic emissions are presented in the above discussions.
In heavy-ion collisions, the created pion source is possibly partially coherent \cite{Glauber:2006gd}.
For a source with partial coherence, the single-pion momentum distribution can be written as the
absolute square of the total amplitude for pion emission,~\cite{CYWong_book}
\begin{equation}
P(\textbf{\pp})=\bigg|{\sum_x}^c A_{\!_C}(\pp)\,e^{ip\cdot x} +{\sum_x}^{\chi}\big|A_{\chi}(x,\pp)\big|\,e^{i\phi_{\chi}(x)} e^{ip\cdot x}\bigg|^2,\\
\end{equation}
where ${\sum}^c$ and ${\sum}^{\chi}$ denote the sums over the coherent and the chaotic source points, respectively.
By expanding the absolute square and casting out the interference terms
with the random chaotic phase $\phi_{\chi}(x)$, which give a negligible contribution, we have
\begin{eqnarray}
P(\textbf{\textit{p}})\!=\!\big|\,{\sum_x}^c\! A_{\!_C}(\pp)\,e^{ip\cdot x}\,\big|^2
\!+\!{\sum_x}^{\chi} \big|A_{\chi}(x,\pp)\big|^2
\!\!=\!P_{\!_C}(\pp)\!+\!P_{\chi}(\pp).
\end{eqnarray}
Because there is no interference effect between the coherent and chaotic emissions in
the single-pion momentum distribution, the total distribution is the sum of the distributions of the
coherent and chaotic sources.

Generally speaking, the particle momentum distribution of an evolving source can be
affected by the source geometry and expansion.  Next, we shall examine the effects
of the source anisotropic geometry and expanding velocity on the pion transverse-momentum
spectrum and elliptic anisotropy for evolving coherent and chaotic sources.

\section{Effects of source anisotropic geometry and expansion on pion transverse-momentum spectrum
and elliptic anisotropy}
To make quantitative comparisons between the effects of coherent and chaotic emissions
on pion transverse-momentum spectrum and elliptic anisotropy, we perform a source parametrization
for both the coherent and chaotic sources.
In relativistic heavy-ion collisions, the source expansion in transverse plane ($xy$ plane)
is anisotropic due to the anisotropic transverse distribution of the energy deposition in
the nuclear overlap zone.
To examine the effects of the source expansion\,(mainly the transverse expansion),
we consider an evolving source with a Gaussian initial spatial distribution in the source
center-of-mass frame (CMF) as
\begin{eqnarray}
\rho_{\rm i-s}(\rr_0)=\frac{(R_xR_yR_z)^{-1}}{\sqrt{(2\pi)^3}}\exp{\left(-\frac{x_0^2}{2R_x^2}
-\frac{y_0^2}{2R_y^2}-\frac{z_0^2}{2R_z^2}
\right)},
\label{GausS}
\end{eqnarray}
where $R_x$, $R_y$, and $R_z$ represent the spatial sizes of the source at an initial time
$t_0$. Then, we assume that each of the source elements has a velocity in the CMF
as \cite{Zhang:2006sw,Yang:2013mxa}
\begin{eqnarray}
v_j(\rr_0)=\textrm{sign}(r_{0j})\cdot a_j\left(\frac{|r_{0j}|}{R_{j,\,max}}\right)^{b_j},
\label{granvelo}
\end{eqnarray}
where $j=x$, $y$, or $z$ denotes the velocity component and ${\rm sign}(r_{0j})=\pm1$
for positive/negative $r_{0j}$, ensuring the source is expansive.  The magnitude of
$v_j$ increases with $|r_{0j}|$, and the rate of increase is decided by the positive
parameters $a_j$, $b_j$, and $R_{j,\,max}$.  In our calculations, we take $R_{j,\,max}=3R_j$,
and consider the source elements with $|\textbf{\textit{v}}|<1$,
which is naturally guaranteed by considering those elements initiated
from the ellipsoidal region $\sum_j{({r_{0j}}/{R_{j,\,max}})^2}<1$,
with the parameters $a_j$ and $b_j$ properly chosen.

The temporal distribution of each source element is parameterized to be Gaussian;
thus, the space-time distribution, $\rho_0(x)$, of a source element initiated from $(t_0,\rr_0)$
can be written in the CMF as $\rho_0(x)=\widetilde{\rho}_0(t-t_0, \,\rr-\rr_0)$, with
\begin{eqnarray}
\widetilde{\rho}_0(x)
=\sqrt{\frac{2}{\pi}}\,\tau_s^{-1}
\exp\left(-\frac{t^2}{2\tau^2_s}\right)\,\delta^{(3)}(~\rr-\textbf{\textit{v}} t~),~~~~(t>0),
\label{rho0CMF}
\end{eqnarray}
where the wide-tilde $\widetilde{\rho}_0(x)$ is the equivalent distribution with the
initial coordinate shifted from $(t_0,\rr_0)$ to $(0,\textbf{{0}})$,
and $\tau_s$ is the duration time of the source element in the CMF.
In the calculations, we consider that all the source elements have the same duration time $\tau_s$ for simplicity.
Furthermore, the velocity $\textbf{\textit{v}}=\textbf{\textit{v}}(\rr_0)$ is in the form of Eq.\,(\ref{granvelo});
thus, $\widetilde{\rho}_0(x)$ is an $\rr_0$-dependent\,(i.e. a source-element-dependent) distribution.

Considering that all the source elements are evolved from the initial source $\rho_{\rm i-s}(\rr_0)$,
we can write the source space-time distribution in the CMF by integrating all the sub-distributions $\rho_0(x)$ together as
\begin{equation}
\rho(x)=\int\! d\rr_0\, \rho_{\rm i-s}(\rr_0)\, \rho_0(x).
\label{rhorho0}
\end{equation}
Note that $\rho_0(x)$ depends on $\rr_0$ as discussed above.
At this point, we have completed the parametrization for both the coherent and chaotic sources.

To calculate the single-pion momentum distribution, we still need to express the pion emission
amplitude for the expanding source.
Generally the amplitude $A(x,\pp)$ for the pion emission at the source point $x$ depends on the
source velocity at $x$,
and can be connected to the amplitude $A'(\pp')$ in the local-rest frame\,(LRF) of the source
element at $x$, as $\sqrt{E_p}A(x,\pp)=\sqrt{E'_p}A'(\pp')$.
Applying the form in Eq.\,(\ref{cohpoint}) as
the LRF amplitude $A'_{_C}(\pp')$ for coherent pion emission,
we have the amplitude in the CMF as
\begin{eqnarray}
\label{cohampCMF}
A_{_C}(\pp)=\frac{\sqrt{E'_p}}{\sqrt{E_p}}\,A'_{_C}(\pp')\!=\!\frac{\sqrt{E'_p}}{\sqrt{E_p}}
\frac{1}{\sqrt{2{E_p}\!\!'(2\pi)^3}}=\frac{1}{\sqrt{2E_p(2\pi)^3}}.~~
\end{eqnarray}
Similarly, from the LRF form in Eq.\,(\ref{P_chS}), we
have $\left|A_{\chi}(x,\pp)\right|^2\!\sim\!\frac{1}{E_p}\frac{(p\cdot u)}{e^{(p\cdot u)/T}\!-\!1}$ for
a chaotic thermal emission,
where $u$ is the CMF 4-velocity of the source element at $x$.  Note that Eq.\,(\ref{cohampCMF})
means the amplitude for coherent emissions, $A_{_C}(\pp)$, is independent of source velocity,
and the form of Eq.\,(\ref{cohpoint}) is appropriate for both static and expanding sources.

The Lorentz invariant momentum distribution, $E_pP(\pp)=dN^3\!/(p_Tdp_Tdyd\phi)$,
for the evolving coherent source can be written according to Eq.\,(\ref{P_c}) as
\begin{eqnarray}
E_pP_{\!_C}(\pp)=E_p\left|\int\! d^4\!x\,e^{ip\cdot x}\,\rho(x)\,A_{_C}(\pp)\,\right|^2,
\label{P_c1}
\end{eqnarray}
where $A_{_C}(\pp)$ is in the form of Eq.\,(\ref{cohampCMF}).
Inserting Eq.\,(\ref{rhorho0}) for the source space-time distribution $\rho(x)$,
one can rewrite the Lorentz invariant momentum distribution with the integral over $\rr_0$ as
\begin{eqnarray}
E_pP_{\!_C}(\pp)=E_p\left|\int\! d\rr_0\,\rho_{\rm i-s}(\rr_0)\,e^{-i\pp\cdot\rr_0}
{\cal A}_{_{C0}}(\rr_0, \pp)\,\right|^2,
\label{P_c2}
\end{eqnarray}
where
\begin{eqnarray}
&&~~~~{\cal A}_{_{C0}}(\rr_0, \pp)=\!\int\! d^4 x~e^{ip\cdot x}\,\widetilde{\rho}_0(\,x\,)\,A_{_C}(\pp)\cr\cr
&&=\!\frac{1}{\sqrt{2E_p(2\pi)^3}}\!\sqrt{\frac{2}{\pi}}\,\gamma_u\tau_s^{-1}\!\!\!
\int_{0}^\infty \!\!dt\, \exp\left[i(p\!\cdot\! u)\,t-\!\frac{t^2}{2(\gamma_u^{-1}\tau_s)^2}\right]~~~~\cr\cr
&&\equiv\,A_{_C}(\pp)\,\,G_0\big[p\!\cdot u(\rr_0)\big],
\label{subamplc}
\end{eqnarray}
with the factor $G_0$ the Fourier transform of the distribution $\widetilde{\rho}_0(x)$.
However, the invariant momentum distribution for the
expanding chaotic source with thermal emissions can be written as
\begin{eqnarray}
&&~~~~E_pP_{\!\chi}(\pp)=E_p\!\int\! d\rr_0\, \rho_{\rm i-s}(\rr_0)\,\left|{\cal A}_{\chi{_0}}(\rr_0, \pp)\right|^2~~~~~~~~~~~~~~~~~~~~~~~~ \cr\cr
&&=\!\int\! d\rr_0\, \rho_{\rm i-s}(\rr_0)\,\frac{p\!\cdot u(\rr_0)}{e^{\left[p\cdot u(\rr_0)\right]/T}\!-\!1}.
\label{P_ch2}
\end{eqnarray}
In Eqs.\,(\ref{subamplc}) and (\ref{P_ch2}), $u=u(\rr_0)=\gamma_u(1,\textbf{\textit{v}}(\rr_0))$
is the 4-velocity of the source element corresponding to the sub-distribution $\rho_0(x)$, with the Lorentz factor $\gamma_u\!=\!(1-\textbf{\textit{v}}^2)^{-1/2}$.
The amplitudes ${\cal A}_{_{C0}}(\rr_0, \pp)$ and ${\cal A}_{\chi_0}(\rr_0, \pp)$ are related to the
pion emissions of the sub-distribution $\rho_0(x)$, for coherent and chaotic sources, respectively.
For the coherent source, it is noted that, due to the interferences in the pion emissions along the (moving-)\,source-element trajectory,
${\cal A}_{_{C0}}(\rr_0, \pp)$ is dependent on the source velocity $u(\rr_0)$,
although the amplitude $\,A_{_C}(\pp)$ related to a point emitter\,[$\delta^{(4)}(x)$] is velocity independent.
For the chaotic source, the momentum distribution $|{\cal A}_{\chi_0}(\rr_0, \pp)|^2$ is in the same form
of $|A_{\chi}(\,x,\,\pp)|^2$,
since there is no interference effect present.
In Eqs.\,(\ref{P_c2}) and (\ref{P_ch2}), the total momentum distributions of coherent and chaotic emissions
can be viewed as the results of the coherent and incoherent superpositions of all the sub-distribution pion emissions, respectively.

\begin{figure}[t]
\includegraphics[scale=0.67]{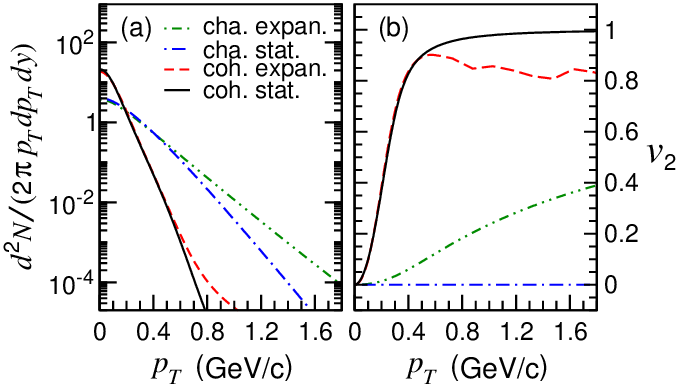}
\caption{(Color online) Pion transverse-momentum spectrum (left-hand panel) and
second-order azimuthal anisotropic coefficient $v_2(p_T)$ (right-hand panel) for
static/expanding coherent and chaotic sources with initial geometric parameters
$R_T=R_z=1$ fm, $S_T=2$, and the duration $\tau_S=3$ fm/$c$.
The velocity parameters for the expanding sources are taken to be $a_x=0.6$, $a_y=0.4$,
$a_z=0.5$, and $b_{x,y,z}=0.5$. Temperature of the chaotic source is 100~MeV. }
\label{sourexpan}
\end{figure}

We plot in Fig.~\ref{sourexpan} the pion transverse-momentum spectrum at central rapidity
and the elliptic anisotropy, $v_2(p_T)=\langle
\,(p_x^2-p_y^2)/p_T^2\,\rangle$, for the static/expanding coherent and chaotic sources.
For both coherent and chaotic sources, the initial transverse and longitudinal spatial size
parameters are taken to be
$R_T\equiv\sqrt{R_xR_y}=R_z=1$~fm, the initial transverse shape parameter is taken to be
$S_T\equiv R_y/R_x=2$, the duration time $\tau_s$ is taken to be 3~fm/$c$, and the
velocity parameters for the expanding sources are taken to be $a_x=0.6$, $a_y=0.4$, $a_z=0.5$,
and $b_{x,y,z}=0.5$, respectively.  In addition, the temperature of the chaotic source is 100~MeV.

It is observed in the left-hand panel of Fig.~\ref{sourexpan} that the source expansion velocity increases
the width of transverse-momentum distribution for chaotic emission\,(e.g. the width $\sqrt{\langle {p_T}^2\rangle}$ for $p_T\!<\!2$\,GeV increases by approximately $16.2\%$),
the effect of which is also referred to as the radial flow.
This expansion velocity effect on the coherent emission is found to be
small\,(e.g., the width increase is approximately $3.7\%$).
Although the ${\cal A}_{_{C0}}(\rr_0, \pp)$ in Eq.\,(\ref{P_c2}) is velocity dependent,
the finally observed momentum distribution that results from the interference effect is close to
that of the static coherent source.
For the static coherent source, ${\cal A}_{_{C0}}(\rr_0, \pp)\!=\!\!\,A_{_C}(\pp)\,G_0(E_p)$ and
is independent of $\rr_0$; thus, we have, from Eq.\,(\ref{P_c2}), that
\begin{eqnarray}
\frac{dN_{^C}^3}{p_Tdp_Tdyd\phi}\bigg|_{\,y=0}
\!\!=E_p \left|{\cal A}_{_{C0}}(\rr_0, \pp)\,\right|^2 \left|\int\! d\rr_0\,\rho_{\rm i-s}(\rr_0)\,e^{-i\pp\cdot\rr_0}\right|^2 \cr\cr\cr
\!\!=\frac{(2\pi)^{-3}}{2}\left|G_0(E_p)\right|^2\exp\left(-R_x^2p_x^2-R_y^2p_y^2\right).~~~
\label{cohstatic}
\end{eqnarray}
The corresponding $v_2$ increases with $p_T$ and approaches to 1 at high
$p_T$ \big[$v_2\!=\!I_1(p_T^2(R_y^2-R_x^2)/2)/I_0(p_T^2(R_y^2-R_x^2)/2)$\big], which can be seen
in the right-hand panel of Fig.~\ref{sourexpan}.
For the expanding sources, the $v_2$ of the coherent emission decreases somewhat in the
$p_T>0.5$~GeV/$c$ region,
but approaches the result of the static source at smaller $p_T$.
At the same time, the anisotropic expansion velocity is more significant for chaotic emission
and leads to the nonzero $v_2$,
which is often referred to as the elliptic flow.

\begin{figure}[t]
\includegraphics[scale=0.67]{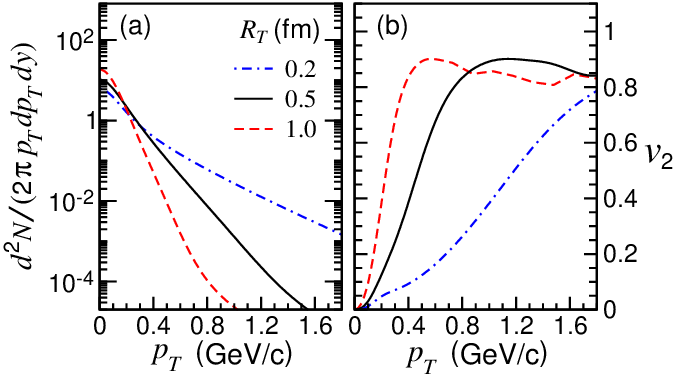}
\caption{(Color online) Pion transverse-momentum spectrum (left-hand panel) and
second-order azimuthal anisotropic coefficient $v_2(p_T)$ (right-hand panel) for
expanding coherent sources with initial geometric parameters $R_T=$ 0.2, 0.5,
and 1.0 fm.  In the calculations, $R_z$ is taken to be the same as $R_T$ and the other
parameters are the same as in Fig. \ref{sourexpan}. }
\label{cohv2specpt}
\end{figure}

Unlike the elliptic flow of the chaotic emission, which is caused by the anisotropic transverse
expansion of the source, the $v_2$ of the coherent emission arises from the initial geometry of
the source already, and is similar as that of the static source, which can be attributed to the
quantum effect\,(related to the interference in single-particle momentum distribution).
To further examine the quantum effect in coherent emission, we present in Fig. \ref{cohv2specpt}
the results of the pion $p_T$ spectrum and $v_2(p_T)$ for the expanding coherent sources
with different initial geometric parameters $R_T=$ 0.2, 0.5, and 1.0 fm.
In the calculations $R_z$ is taken to be the same as $R_T$ and the other parameters are the same
as in Fig.~\ref{sourexpan}.
One can observe that the transverse-momentum spectrum is "harder" for the source with a smaller
$R_T$, while the increase of $v_2$ with $p_T$ covers a larger range for smaller $R_T$.
These effects are also exposed in the analytical expressions for the static coherent
source\,[Eq.\,(\ref{cohstatic}) and the corresponding expression for $v_2$].
In addition, we check the results with different source expansion parameters,
and find that within the current framework both the $p_T$ spectrum and $v_2(p_T)$ show substantially more
sensitivities to the source initial geometry than to the expansion parameters\,(flow effects).

\section{Results of partially coherent source}
As seen in the preceding section, the transverse-momentum spectrum and $v_2$ of
the coherent emission are mainly attributed to the quantum effect
and are sensitive to the initial geometry of the coherent source.
They are different from those of the chaotic source, which are significantly affected by the source
dynamical evolution.
In high-energy heavy-ion collisions, the created pion source is possibly partially coherent \cite{Glauber:2006gd}.
The significant suppression of multi-pion Bose-Einstein correlations recently
observed by the ALICE Collaboration in the Pb-Pb collisions at the LHC
\cite{ALICE-HBT14,ALICE-HBT16} indicates that the pion emission may have a considerable
degree of coherence.
In this section, we investigate the pion transverse-momentum
spectrum and elliptic anisotropy in a partially coherent source model constructed as follows:

(1) For the chaotic part, we consider it a hydrodynamically evolving source.
Note that it is implied in most of the conventional hydrodynamics models that the source is chaotic.
A glimpse of this can be seen in the commonly used Cooper-Frye procedure\,\cite{Cooper:1974mv},
in which the total spectrum of the decoupled particles is the summation of the
spectra of the particles decoupled\,(emitted) at different space-time positions.
In this work, we adopt the viscous hydrodynamics code in the iEBE-VISHNU code package\,\cite{Shen:2014vra},
which is widely used in the study of relativistic heavy-ion collisions.
To be specific, we use the MC-KLN model based on the ideas of parton
%
%
saturation in the CGC\,\cite{Kharzeev:2001yq,Kharzeev:2004if} for the initial condition.
For one studied centrality class,
we use a smooth, single-shot initial condition\,\cite{Qiu:2011iv,Shen:2014vra}, obtained by
averaging 1000 randomly generated events with the maximum specified eccentricity, as the input
of the hydrodynamical evolution.  Since at this step we do not focus on the effects of the
event-by-event fluctuations, the single-shot event, which carries the main feature of the
chaotic source, is economical for our simulation.
With the initial condition prepared, we perform the (2+1)-dimensional viscous hydrodynamics
code VISHNew, an improved version of VISH2+1\,\cite{Song:2007ux,Song:2008si,Heinz:2008qm},
incorporating the lattice QCD-based equation of state s95p-PCE\,\cite{Huovinen:2009yb}, to
evolve the source.  We set the initial time of the hydrodynamical evolution at $\tau_0=0.6$ fm/$c$
and the decoupling temperature at $T_{\rm dec}=120$ MeV, which is the same as used in Ref.\,\cite{Shen:2011eg}.
After the Cooper-Frye procedure\,\cite{Cooper:1974mv}, we obtain the pion
momentum distribution for the hydrodynamic (chaotic) source by utilizing the AZHYDRO resonance
decay code \cite{Sollfrank:1990qz,Sollfrank:1991xm,Sollfrank:1996hd}.

(2) For the coherent part, since the real mechanism responsible for the coherent pion
production is not yet clear, we adopt the parameterized expanding source as introduced
in the preceding section.  The initial source spatial distribution is a Gaussian one as in Eq.
(\ref{GausS}), the source expansion velocity is described by Eq. (\ref{granvelo}), and the Gaussian
temporal distribution is given by Eq.\,(\ref{rho0CMF}).

(3) For simplicity, we considered an idealization that the chaotic and coherent sources evolve
independently in the partially coherent source model.
Moreover, assuming that the formations of the chaotic and coherent parts are spatially correlated,
we consider an ideal case in which their spatial distributions in the transverse plane are oriented in the
same direction, i.e., their initial eccentricities are maximized with respect to the same reference
plane\,("participant plane").
This is similar as the case of a trapped atomic gas in a Bose-Einstein condensate\,\cite{Anderson:1995gf,Davis:1995pg}.

\begin{figure*}[htb]
\includegraphics[scale=0.6]{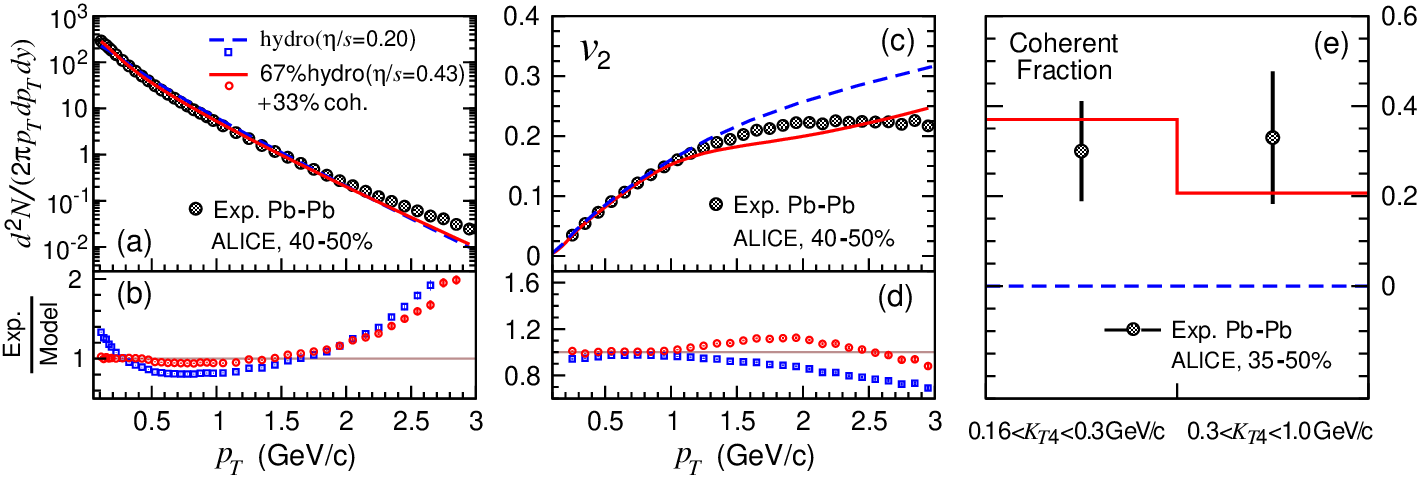}
\caption{(Color online) Comparison of results of pion transverse-momentum spectrum
(left-hand panels), elliptic anisotropy (middle panels), and coherent fraction (right-hand panel) of
the hydrodynamical chaotic source and the partially coherent source with experimental
data of pion transverse-momentum spectrum \cite{ALICE-spe13}, elliptic anisotropy
\cite{ALICE-v2-15}, and the coherent fraction in the two $K_{T4}$ regions, $0.16<K_{T4}
<0.3$~GeV/$c$ and $0.3<K_{T4}<1$~GeV/$c$ \cite{ALICE-HBT16}, measured by the ALICE
Collaboration in Pb-Pb collisions at $\sqrt{s_{NN}}=2.76$~TeV and in the centrality
regions of 40--50\% (left and middle panels) and 35--50\% (right panel).
Error bars of experimental data of pion transverse-momentum spectrum and elliptic
anisotropy are smaller than symbol size and are difficult to observe.  Quantity $K_{T4}$
is the four-pion average transverse momentum defined as $K_{T4}=\frac{1}{4}\big|\pp_{T,1}
+\pp_{T,2}+\pp_{T,3}+\pp_{T,4}\big|$. }
\label{parcohptv2}
\end{figure*}

From the measurements of four-pion Bose-Einstein correlation functions in the Pb-Pb collisions
at $\sqrt{s_{NN}}=2.76$~TeV \cite{ALICE-HBT16}, the coherent fraction is approximately 30\% and has
no obvious centrality dependence.
On the other hand, the observed suppression on the correlations
extends at least up to $p_T\!\sim\!340$ MeV/$c$ \cite{ALICE-HBT16}.
Accordingly, we estimate that the initial (minimum) transverse size of the coherent emission
region is $\lesssim \hbar/(2\times340$~MeV/$c$)$=0.29$~fm.

In the left-hand and middle panels of Fig.~\ref{parcohptv2}, we show the results of the
pion $p_T$ spectrum and $v_2(p_T)$, respectively, for the Pb-Pb collisions at $\sqrt{s_{NN}}=
2.76$~TeV in the 40--50\% centrality region.
The experimental data\,(black bullets) measured by the ALICE Collaboration~\cite{ALICE-spe13,ALICE-v2-15}
are shown for comparison.  Here, the blue dashed lines and squares are the
results of the conventional pure hydrodynamical chaotic source with the ratio of shear viscosity to entropy density
$\eta/s=0.20$ \cite{Shen:2011eg}.  The red solid lines and circles are the results of the
partially coherent source model constructed with the hydrodynamical chaotic part with $\eta/s=0.43$
and the coherent part with the initial geometry parameter $R_T=R_z=0.25$~fm and $S_T=2$.
The duration time and expansion velocity parameters for the coherent source are taken to be $\tau_s=2$~fm,
$a_x=0.4$, $a_y=0.3$, and $a_z=0.35$, and the other parameters are the same as in the preceding section.
For the partially coherent source, the total fractions of chaotic and coherent emissions,
obtained with the particle yields in the transverse-momentum range $p_T<3$~GeV/$c$, are 67\% and 33\%,
respectively. We also show in the left-hand- and middle-bottom panels of Fig. \ref{parcohptv2} the ratios
of the experimental data to the results of the models.
One can observe that the results of the transverse-momentum spectrum and elliptic anisotropy
of the partially coherent source are slightly more consistent with the experimental data than
those of the pure hydrodynamical chaotic source.

\begin{figure*}[htb]
\includegraphics[scale=0.6]{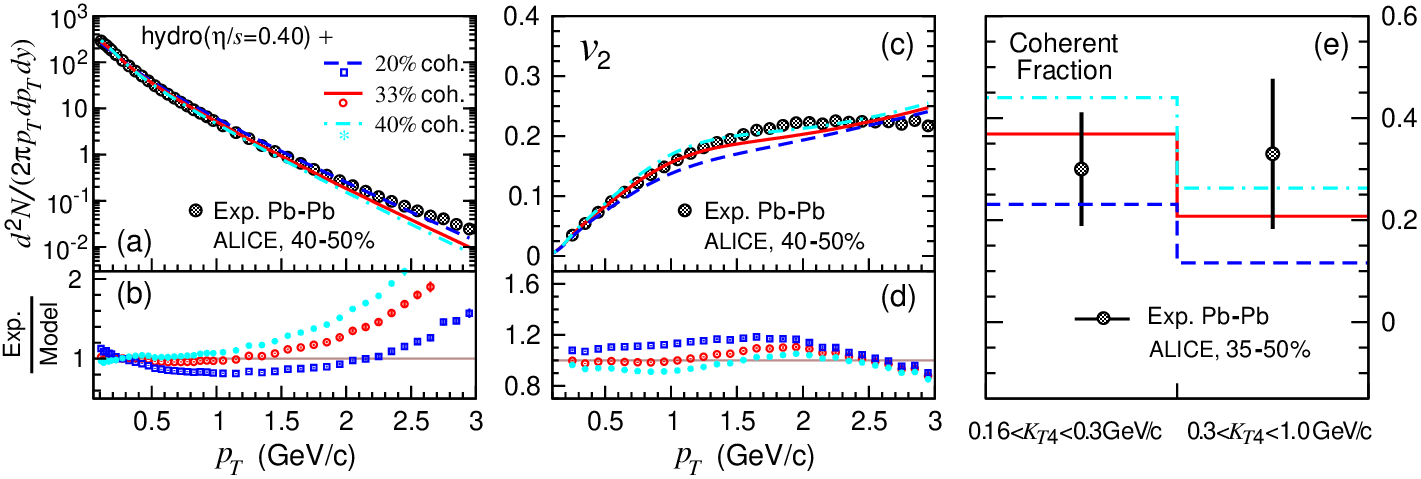}
\caption{(Color online) Comparison of results of pion transverse-momentum spectrum
(left-hand panels), elliptic anisotropy (middle panels), and coherent fraction (right-hand panel) of
partially coherent sources, which have different total fractions of coherent emission,
with experimental data the same as in Fig.\,\ref{parcohptv2}. }
\label{parcohptv2_3}
\end{figure*}

In the right-hand panel of Fig.~\ref{parcohptv2} we show the coherent fraction of the partially
coherent source and the results extracted from the experimental measurements of four-pion
Bose-Einstein correlations in two $K_{T4}$ regions, $0.16<K_{T4}<0.3$~GeV and $0.3<K_{T4}
<1.0$~GeV, in the Pb-Pb collisions at $\sqrt{s_{NN}}=2.76$~TeV and in the 35--50\% centrality
region \cite{ALICE-HBT16}, where $K_{T4}$ is the four-pion average transverse momentum defined
as $K_{T4}=\frac{1}{4}\big|\pp_{T,1}+\pp_{T,2}+\pp_{T,3}+\pp_{T,4}\big|$.  The coherent fraction
of the pure hydrodynamic source is zero.  However, the results of the partially coherent source are
consistent with the experimental data.
For the partially coherent source, the transverse-momentum distribution of the
coherent part decreases with $p_T$ faster than that of the chaotic part\,(similar as seen in Fig.\,\ref{sourexpan});
thus, the coherent fraction decreases with the transverse momentum.

We further investigate the variations of the results of the partially coherent source model
with different total fractions of coherent emission, for the same observables as in
Fig.~\ref{parcohptv2}.  Here, $\eta/s$ for the chaotic part is taken to be $0.4$.
The experimental data presented in Fig.~\ref{parcohptv2_3} are the same as in Fig.~\ref{parcohptv2}.
The results of the pion $p_T$ spectrum and $v_2(p_T)$ are found to be sensitive to the total fraction
of coherent emission, mainly due to the fact that the coherent part tends to produce a relatively steep spectrum
and a larger $v_2(p_T)$.
The results corresponding to the total fraction of coherent emission, 33\%, are more consistent
with the experimental data of the pion $p_T$ spectrum and $v_2(p_T)$, and this
fraction of coherent emission is consistent with the experimental analysis results of four-pion Bose-Einstein correlations.

It is noted that, with the influence of the coherent emission taken into account,
the specific shear viscosity of the hydrodynamical chaotic source in this model
is taken to be larger, $\sim0.4-0.43$, relative to 0.2 \cite{Shen:2011eg},
to better describe the experimental data.
However, more systematic studies on source coherence and viscosity should be done to make more
definitive statements about the viscosity of partially coherent sources.

\section{Summary and discussion}
\label{conclusion}
Motivated by the recent experimental observation of the suppression of multi-pion Bose-Einstein
correlations in Pb-Pb collisions at $\sqrt{s_{NN}}=2.76$~TeV at the LHC \cite{ALICE-HBT14,ALICE-HBT16}, we have studied coherent pion emission and its influences on the pion transverse-momentum spectrum and elliptic anisotropy in relativistic heavy-ion collisions.
We have constructed a partially coherent source by combining the chaotic emission source evolving
with viscous hydrodynamics and a parameterized coherent emission source.  It is found that the
influences of coherent emission on the pion transverse-momentum spectrum and elliptic anisotropy are
related to the initial size and shape of the coherent source,
largely due to the interference effect in the single-particle momentum distribution.
However, the effect of the source dynamical evolution on coherent emission is relatively small.
The pion $p_T$ spectrum and $v_2(p_T)$ generated by the partially coherent source
with a total fraction of coherent emission, 33\%, reproduced the experimental data\,\cite{ALICE-spe13,ALICE-v2-15}
in Pb-Pb collisions with 40--50\% centrality at the LHC.
This coherent fraction is consistent with the experimental analysis results
of four-pion Bose-Einstein correlations\,\cite{ALICE-HBT16}.
In addition, the small initial transverse radius
of the coherent source, $R_T=0.25$~fm, is consistent with the experimental observation, specifically that
``~The suppression observed in this analysis appears to extend at least up to $p_T\sim340$~MeV/c"
\cite{ALICE-HBT16}, which may provide some clues of the origin of the coherence.

It is usually implied that the particle source is chaotic in the study of relativistic heavy-ion
collisions, which has indeed captured the main feature of particle emission.
However, there have been some indications of coherent particle emission, although its mechanism
is not yet clear.
The partially coherent source model presented in this letter reveals some interesting effects of
coherent emission on the transverse-momentum spectrum and elliptic anisotropy.
Additional work remains to systematically study the effect of coherent emission
on experimental observables and to understand the mechanisms of coherent emission in relativistic
heavy-ion collisions.

\begin{acknowledgments}
This research was supported by the National Natural Science Foundation of China under Grant Nos.
11675034 and 11275037.
\end{acknowledgments}

\end{document}